\documentstyle[12pt,a4]{article}

\def\lesssim{\mathrel{\mathpalette\vereq<}}
\def\gtrsim{\mathrel{\mathpalette\vereq>}}
\makeatletter
\def\vereq#1#2{\lower3pt\vbox{\baselineskip1.5pt \lineskip1.5pt
\ialign{$\m@th#1\hfill##\hfil$\crcr#2\crcr\sim\crcr}}}
\makeatother

\begin{document}

\begin{titlepage}
\begin{center}
~{} \hfill    CERN-TH/99-218\\
~{} \hfill LBNL-44140 \\
~{} \hfill UCB-PTH-99/30\\

\vskip .1in

{\large \bf The Small Observed Baryon Asymmetry\\
from a Large Lepton Asymmetry}%
\footnote{HM was supported in part by the U.S.
Department of Energy under Contracts DE-AC03-76SF00098, in part by the
National Science Foundation under grant PHY-95-14797.  Both JMR and
HM thank the Alfred P. Sloan Foundation for support.}

\vskip 0.3in

John March-Russell$^a$, Hitoshi Murayama$^{b,c}$,
and Antonio Riotto$^a$

\vskip 0.05in

{\em $^a$Theory Division\\CERN, CH-1211\\Geneva 23, Switzerland}

\vskip 0.05in

{\em $^b$Department of Physics\\ University of California\\
Berkeley, CA 94720, USA}

\vskip 0.05in

{\em $^c$Theoretical Physics Group\\ Lawrence Berkeley National 
Laboratory\\ University of California\\
Berkeley, CA 94720, USA}

\vskip 0.05in

\end{center}

\vskip .1in

\begin{abstract}
Primordial Big-Bang Nucleosynthesis (BBN) tightly constrains 
the existence of any additional relativistic degrees of freedom
at that epoch.  However a large asymmetry in electron neutrino number
shifts the chemical equilibrium between the neutron and proton at
neutron freeze-out and allows such additional particle species.
Moreover, the BBN itself may also prefer such an asymmetry to
reconcile predicted element abundances and observations. 
However, such a large asymmetry appears to be in conflict with
the observed small baryon asymmetry if they are in sphaleron mediated
equilibrium.  In this paper we point out the surprising fact that in the
Standard Model, if the asymmetries in the electron number and the
muon number are equal (and opposite) and of the size required to
reconcile BBN theory with observations, a baryon asymmetry of the
Universe of the correct magnitude and sign is automatically generated
within a factor of two.  This small remaining
discrepancy is naturally remedied in the supersymmetric Standard
Model.  

\end{abstract}

\end{titlepage}

\newpage

\section{Introduction}

Primordial Big Bang Nucleosynthesis (BBN) is without doubt one
of the biggest
successes of early universe cosmology.  Not only does it provide a 
stringent test of the Big Bang model, predicting the light element 
abundences as a function of only a single parameter, $\eta=n_b/n_\gamma$,
the cosmological baryon to photon ratio, but it also supplies important
constraints on particle physics, the most well-known example being the
determination of the number of light neutrino species.  Given the
consistency between the primordial abundance of light elements
(inferred from observation extrapolated back to the primordial values)
and theoretical calculations, BBN does not leave much room for extra
particles which otherwise could have existed in the early universe.  

Many extensions of physics beyond the Standard Model (SM), however,
introduce additional relativistic degrees of freedom at the epoch of
BBN.  A small selection of such new
light degrees of freedom include: one or more sterile neutrinos which
might be required by the neutrino oscillation data; a light gravitino
as a consequence
of a low fundamental scale of supersymmetry breaking as in the
gauge-mediated scenarios; a hadronic axion in the hot dark matter
window, and many other examples.  If such new light degrees
of freedom exist,
the expansion rate at the BBN epoch is faster, resulting in an
earlier freeze-out of neutrons and hence a larger number of them,
therefore overproducing $^4$He.  Taking the BBN constraint
seriously, it is then necessary to modify standard BBN, and the
simplest and most elegant possibility is a large lepton
asymmetry\footnote{One other possibility discussed in the literature
is that of a late-decaying $\nu_{\tau}$.}, a possibility which is not ruled
out by current observational limits \cite{ks92,us,pastor}.
Specifically, a large positive asymmetry
in the electron number implies an excess in the number of electron
neutrinos over that of electron anti-neutrinos, thereby shifting the
chemical equilibrium between protons and neutrons towards protons.
This results in a smaller number density of neutrons after the
freeze-out and hence in a reduced $^4$He abundance.  This effect can
therefore compensate the effect of the larger expansion rate due to
the additional particle species.

Moreover in recent years, with the advent of new and refined data on
the relative abundances of the light elements, there may be appearing a slight
but significant discrepancy between the data and the theoretical
predictions.  In particular, if
the recent low measurements~\cite{lowD} of the primordial Deuterium
abundance are correct and the $^4$He abundance is as low as reported
in \cite{Olive}, then some modification of the standard BBN scenario seems
to be required independent of the conjectured existence of new light degrees
of freedom.  The most promising such modification is again the
assumption of a positive chemical potential for electron neutrinos
which reduces the final $^4$He abundance closer to the reported
value.  It is noteworthy that
the preferred sign of the electron asymmetry is the same for both
purposes: to compensate the effect of additional particle species and
to bring the BBN prediction closer to observations.
Of course, given the uncertainties in the data, it is not
clear if this is really required by primordial nucleosynthesis.  It
is, however, useful to explore such modifications of the standard
Big-Bang scenario to see if they are either disfavored by other data,
or serve some further, unexpected, purpose.  

On the other hand, there is an apparent contradiction of an
assumption of large lepton asymmetry with the very small observed
baryon asymmetry.  This arises
from the presence of sphaleron mediated transitions at temperatures of
the weak scale and above which tend to quickly equilibriate the lepton
and baryon asymmetries, resulting in far too large a baryon asymmetry
today.  There are three logical possibilities for how a large $\nu_e$
lepton asymmetry can be compatible with the small baryon asymmetry:
(1) Sphalerons were never in equilibrium, (2) The lepton asymmetry is
generated after the electro-weak phase transition but before BBN, and,
(3) The {\it total} lepton asymmetry across all three generations is
zero.

In this letter we focus on the third possibility -- in particular the
case where $L_e=-L_\mu\neq 0$ and $L_\tau=0$ -- and show, in
Section~3, that it has a very pretty and unexpected consequence -- the
natural generation within the {\it Standard Model} of a small baryon
asymmetry of the observed size, at least within a factor of two, and
with the correct sign!  This
numerical coincidence is quite remarkable, especially given the
simplicity and naturalness of the baryon asymmetry generation
mechanism.  The fundamental reason for the small baryon asymmetry in
this case, $L_e=-L_\mu\neq 0$, is quite simple; it is just a
consequence of the small muon Yukawa coupling.  As we show in
Section~4, if one goes to the minimal supersymmetric standard model
(MSSM) then even the factor of two discrepancy between the predicted and
observed baryon asymmetry disappears for large $\tan(\beta)$.
(Section~2 contains a more extensive discussion of the reasons
for considering a Lepton asymmetry, together with its
possible size and sign.)

\section{BBN with Large Lepton Number}

We will now argue in detail that it is useful to explore the possibility that
there may be a slight modification of standard BBN, and that such
modifications are certainly not disallowed and are possibly even
favored by the light element abundances.  

Many particle physics models beyond the SM introduce additional
particle species which could be relativistic and thermal at the BBN
epoch.  Probably the most discussed such example is a sterile neutrino
(or many of them, especially in the
context of neutrinos from large extra dimensions \cite{largeDnu}).
If one takes all existent hints for neutrino
oscillations seriously, namely the atmospheric neutrino oscillations,
solar neutrino deficit and the results from the LSND experiment, the
data cannot be accommodated by neutrino oscillations between the three
known species: $\nu_e$, $\nu_\mu$, $\nu_\tau$.  The reason is simple. The
three hints for oscillations listed above require different values of
the mass-squared differences $\Delta m^2$, and with three neutrinos only,
the sum of $\Delta m^2$ should vanish.  The only known way to explain
the data fully by neutrino oscillations is by introducing an
additional ``sterile neutrino'' $\nu_s$, thereby allowing yet another
mass-squared difference to account for three oscillation modes.
However, neutrino oscillations should have occurred in the early universe
as well, thus producing sterile neutrino states.  In order not to
overproduce $^4$He due to the additional sterile
neutrino energy density, the quoted bounds are \cite{sterilebounds,FVPRL}
\begin{eqnarray}
        & & \Delta m^2 \sin^4 2\theta 
        \lesssim 5 \times 10^{-6} {\rm eV}^2, \qquad \nu = \nu_e,
        \nonumber \\
        & & \Delta m^2 \sin^4 2\theta 
        \lesssim 3 \times 10^{-6} {\rm eV}^2, \qquad \nu = \nu_{\mu,\tau}.
\end{eqnarray}
These constraints, taken literally, imply that sterile neutrinos
cannot be responsible for atmospheric neutrino oscillations or the large
angle MSW solution to the solar neutrino problem.  The existence of a
sterile neutrino exceeding the above bounds would increase the
effective number of neutrinos at BBN by one: $\Delta N_\nu = 1$.

In supersymmetric theories, a light gravitino $\tilde{G}$ may be
present at the BBN epoch as well.  According to the estimate in
Ref.~\cite{MMY}, the gravitinos remain thermal down to the BBN epoch if
\begin{equation}
        m_{3/2} \lesssim 10^{-13}{\rm GeV} 
        \left(\frac{m_{\tilde{l}}}{100 \rm GeV}\right),
\end{equation}
due to the process $l^+ l^- \rightarrow \tilde{G} \tilde{G}$.  This
roughly corresponds to a primordial supersymmetry breaking scale
below a TeV.  Such a low scale is not expected in the conventional
hidden sector models or gauge mediation, but can occur in models where
the supersymmetric standard model is directly involved in the
mechanism of dynamical supersymmetry breaking (see, {\it e.g.}, the
model in Ref.~\cite{ALT}).  Because the produced gravitino states are
dominantly helicity $\pm 1/2$ (the would-be Nambu-Goldstino state), they
increase the effective number of neutrinos by $\Delta N_\nu = 1$.

Invisible axions are another candidate particle that
could be present at the BBN epoch.  Despite
strong constraints from astrophysics, a hadronic (KSVZ) axion in the
mass range 3--20~eV is allowed as long as its coupling to the
photon is accidentally suppressed \cite{axionHDM}.  This is an
interesting window for a Hot Dark Matter component of the universe
which some recent analyzes of large scale structure prefer \cite{GS}
(however, for conflicting views, see \cite{noHDM}).  The axion
in this mass window would
contribute to the energy density as an equivalent of $\Delta N_\nu
=0.4$--0.5 \cite{CC} and is marginal from the BBN point of view.

Yet another example of an exotic particle which might
be in thermal contact during BBN is represented by the majoron,
the Goldstone boson associated to the spontaneous breakdown of
lepton number.  Majorons stay in thermal equilibrium as long
as $\tau$-neutrinos, and provide a contribution to 
$\Delta N_\nu$ of about 0.6 \cite{maj}. 

Given these important constraints from BBN on particle physics models,
it is important to ask how rigid the constraint actually is.  
In this regard it is interesting to note that
the BBN itself may require some modifications.

Specifically, if one takes
the low Deuterium measurement \cite{lowD} and the reported statistical
average of the $^4$He abundance extrapolated to zero metalicity
\cite{He}, they cannot be reconciled with detailed BBN calculations
by choosing an appropriate value of $\eta$, the baryon to photon ratio.  Of
course, it is not yet established that these measurements are
reliable.  For instance, one should take seriously the conflicting
measurement of the Deuterium abundance based on the same technique
which returns a high value \cite{highD}, even though it has been
challenged on the basis of a possible overlap with a foreground cloud
and less systematic checks than the low abundance observation. 
(It is interesting to note that by including turbulence effects
in the extraction of the D/He ratio~\cite{turb}, all the data is
cosistent with a low value of D/He $\simeq 3.5 - 5.2 \times 10^{-5}$.)
The ``best'' determination of the $^4$He abundance has also been
challenged by a re-analysis of the more-or-less the same data set
\cite{Russian}.  Nevertheless there is motivation for considering
modifications to BBN which can reconcile the ``best'' determinations
of element abundances.  Most certainly, such a modification is allowed
by current data.  (For a recent review see Ref.~\cite{sarkar}.)  

It is noteworthy that {\it both} the presence of additional
relativistic degrees of freedom and the apparent inconsistency between
the D and $^4$He abundances prefer a mechanism to reduce the effective
number of neutrinos $N_\nu$. 
Two such possibilities have been proposed in the
literature: 
\begin{enumerate} 
\item A late-decaying $\nu_\tau$ with a mass of $m_{\nu_\tau} \sim
10$~MeV and a lifetime of $\tau \sim 10^{-2}$--$1$~sec
\cite{Kawasaki,KKS1}. 
\item A large chemical potential for $\nu_e$ \cite{KKS2}.
\end{enumerate}
The former proposal is interesting from the collider physics point of
view because it is testable in the forthcoming $B$-factory experiments
\cite{FMMS}.

In this letter we focus on the second possibility.  Here the idea is that the
presence of a large chemical potential for $\nu_e$ makes the $\nu_e$ number
density larger than the thermal number density without
chemical potential, which in turn changes the chemical equilibrium of
the reaction
$\nu_e n \leftrightarrow e^- p$ etc.  The presence of a positive
chemical potential for $\nu_e$ shifts the equilibrium towards the
right-hand side, which reduces the neutron number density at the
freeze-out.  Therefore the $^4$He abundance is reduced for a given
value of $\eta$.  Since the $D$ abundance \cite{lowD} prefers a
relatively large value of $\eta$, which prefers a large
$^4$He abundance, the reduced prediction for the $^4$He abundance
would allow additional relativistic degrees of freedom present at the
BBN epoch or reconciles the apparent conflict between the observations
and the calculations.\footnote{In the case of neutrino oscillations to 
a sterile neutrino, the interplay between the neutrino oscillations 
and thermalization can be quite complicated \cite{sterilecomplications}.  
However, a large primordial lepton asymmetry which exists from the pre-BBN 
era does persist \cite{FVPRL} and can allow the sterile neutrinos.  
This differs from the situations discussed in \cite{sterilecomplications} 
where the lepton asymmetry was assumed to vanish primordially ({\it 
i.e.}\/, before the BBN era).}

The electron-neutrino chemical potential affects the neutron-to-proton
ratio at the freeze-out as 
\begin{equation}
        \left( \frac{n}{p} \right)_{\xi_{\nu_e}\neq 0}
        = \left( \frac{n}{p} \right)_{\xi_{\nu_e}= 0} e^{-\xi_{\nu_e}},
\end{equation}
where $\xi_{\nu_e} = \mu_{\nu_e}/T$ at the freeze-out temperature.
The effect of the extra degrees of freedom on $^4$He abundance is
given by an analytic fit \cite{KT}: $\Delta Y_P = 0.0075 \Delta g_* =
0.013 \Delta N_\nu$.  Therefore, an approximate dependence of $Y_P$ on
the extra degrees of freedom and the chemical potential is given by
\begin{equation}
        Y_P = \left(0.225 + 0.025 \log_{10} \left(
        \frac{\eta}{10^{-10}} \right) 
                + 0.013 \Delta N_\nu \right) e^{ - \xi_{\nu_e}}
\end{equation}
for $\tau_{1/2} (n) = 10.24$~minutes.  The low D measurement requires
$\eta \simeq 5 \times 10^{-10}$ and hence $Y_P \simeq 0.242$
which is beyond the quoted $Y_P = 0.234 \pm 0.002 \pm 0.005$ \cite{He}
(see, however, a conflicting number $Y_P = 0.244 \pm 0.002 \pm 0.005$
\cite{Russian}).  This would require $\xi_{\nu_e} \sim 0.0034$.  This
approximate discussion also tells us that an additional degrees of
freedom with $\Delta N_\nu =1$ can be compensated by $\xi_{\nu_e} =
0.056$. 

The size of the chemical potential favored to reconcile the
observations and the BBN calculations of the light element abundances
were studied by intensive numerical analysis in  Ref.~\cite{KKS2}.  The
result is $\xi_{\nu_e} = (4.3 
\pm 4.0) \times 10^{-2}$ at 95\% CL, quite close to the rough estimate
given above.  From this the electron-number per
photon ratio is given by
\begin{equation}
   \frac{n_{\nu_e} - n_{\bar{\nu}_e}}{n_\gamma} =
  \frac{\pi^3}{12 \zeta(3)} \left( \frac{T_{\nu_e}}{T_\gamma}
  \right)^3
  \left( \frac{\xi_{\nu_e}}{\pi} \right) + {\cal O}(\xi^3).
\end{equation}
Since $T_{\nu_e} = T_\gamma$ in the relevant temperature regime, and
the total entropy density is $s = \frac{43}{4} \frac{2\pi^2}{45} T^3$ (from
photons, electrons, positrons and three neutrinos), we find the ``preferred"
electron-number to entropy ratio $L_e$  to be
\begin{equation}
  L_e^{{\rm NUC}} = \frac{15}{43}
  \frac{\xi_{\nu_e}}{\pi^2} = (1.52 \pm 1.41) \times 10^{-3}.
\label{eq:inputLS}
\end{equation}
For the purpose of allowing 
an extra relativistic degree of freedom at the epoch of BBN, we would also
require an additional contribution the electron-number to entropy ratio
of this same magnitude and sign.  Thus we take
\begin{equation}
	L^{\star}_e\sim 2L_e^{{\rm NUC}} = (3.04 \pm 2.82) \times 10^{-3}
\end{equation}
as the  favored value of the lepton asymmetry
both by compensating an additional relativistic degree of freedom at
the BBN epoch and by reconciling the discrepancy between the theory
and observation in the BBN itself.

\section{Small Baryon Number from Large Lepton Number}

The most uncomfortable aspect of a large chemical potential for $\nu_e$
is the consistency with the small observed baryon asymmetry.  An almost
universal theoretical
prejudice is that the baryon asymmetry is a consequence of
non-trivial dynamics in the Early Universe, with the three Sakharov
conditions being met: (1) the existence of a baryon-number violating
interaction, (2) departure from thermal equilibrium, and (3)
CP-violation.  If there were also a chemical potential for $\nu_e$, or
in other words, an asymmetry in the electron number, it should also
be a consequence of similar dynamics in the Early Universe.  It then
appears unnatural that the lepton asymmetry is many orders of
magnitude larger than the baryon asymmetry if they are generated by
similar mechanisms.

The uncomfortableness mentioned above becomes a conflict in the view
of the following consideration.  Given the difficulty in generating a
large enough baryon asymmetry purely from the electroweak phase transition,
the much larger preferred size of the lepton asymmetry from the BBN, Eqn.~(\ref{eq:inputLS}),
is highly unlikely to be a consequence of physics at or below the electroweak
scale.  However, above the electroweak phase transition, neither baryon- or
lepton-number is conserved, but only $B-L$ because of sphaleron mediated
transitions and the electroweak $B$ and $L$ anomalies~\cite{KRS,AKY,AM,KS}.
Furthermore, the chemical equilibrium induced by
sphaleron transitions enforces the baryon- and lepton-asymmetries to be
of the same orders of magnitude.

There are three logical possibilities to avoid this conflict:
\begin{enumerate}
\item The large lepton asymmetry is generated below the electroweak
  scale.
\item The sphaleron transition was never in equilibrium below the
  temperature at which the lepton asymmetry was generated.
\item The total lepton asymmetry vanishes, while the individual
  lepton-flavor asymmetries do not.
\end{enumerate}
We already argued that the first possibility is unlikely, even though
it is logically possible.  The second possibility arises if the large
lepton number asymmetry causes a Bose condensate of
electroweak-doublet scalar fields \cite{BBD,Linde,DMO,RS}.  In the
Standard Model the preferred value of the lepton
asymmetry from nucleosynthesis considerations is below
the critical value \cite{BRS} at which the Higgs
doublet acquires a large expectation value and thus
at temperatures above the
electroweak scale the sphaleron transition is still in equilibrium.
The same is true in the case of the MSSM as recently shown
in Ref.~\cite{bo}.  Note, in 
particular,  that if the squark and slepton masses are heavier than the
electroweak phase transition temperature of 100--200~GeV, they are
irrelevant to this discussion and the situation is the same as in the
SM and hence the sphaleron transitions are active.  Moreover, even if one
manages to keep sphaleron transitions out of equilibrium, it still
does not resolve the question why the lepton asymmetry is so much
larger than the baryon asymmetry.  From these considerations, we find
the third possibility to be the most interesting one, which has not been
discussed in the literature so far.

The baryon and the lepton asymmetries are determined by the $B-L$
asymmetry via sphaleron-induced chemical equilibrium.
For the Standard Model \cite{KS,HT}:
\begin{equation}
  B = \frac{8N_G + 4N_H}{22N_G + 13N_H} (B-L),
\end{equation}
where $N_G=3$ is the number of generations and $N_H$ is the number of
Higgs doublets (1 in the SM).  In the presence of
the supersymmetric particles, the formula is slightly modified
\cite{IIMS}.  Therefore, if the total lepton asymmetry vanishes, the
total baryon number also vanishes.  This way, one can obtain a
vanishing baryon asymmetry even in the presence of individual
flavor-dependent lepton asymmetries.

The above formula is usually assumed to hold above the electroweak phase
transition temperature, while it requires modification after the phase
transition because of finite mass effects.  However, even above the phase
transition temperature, the effects of thermal masses need to be
considered.  Such effects are small and usually ignored, but they cannot be
ignored in the presence of the large individual lepton numbers of
interest in this letter.

The final resulting baryon asymmetry depends on when the sphaleron
transition freezes out, which in turn depends on whether the electroweak
phase transition is strongly first-order or not \cite{review}.
Given the experimental lower bound on the Higgs mass of
about 95 GeV together with the results of current large-scale
numerical lattice simulations~\cite{lattice}
and analytic arguments~\cite{ewpt},
the phase transition in the Standard Model is certainly not a strongly
first order transition, while in the case of the MSSM a weakly first-order
transition or smoother is favored over much of the parameter space.
In the case that the phase transition is second order,
or if the sphalerons are still active after a first order phase
transition ({\it i.e.}, a weakly first-order transition with $\langle
\phi(T)\rangle/T \leq 1$, being $\langle\phi\rangle$ the 
vacuum expectation value of the Higgs field),
there are two contributions to the resulting baryon asymmetry.
These flavor-dependent effects both arise from the interaction
of electrons and muons with the Higgs boson via
their Yukawa couplings.  (The two effects correspond to the
interactions with condensed and real Higgs bosons respectively.)
The total flavor-dependent effect was
estimated in Ref.~\cite{KS}, and in the
case of vanishing total lepton asymmetry $L_e + L_\mu=0$, we find
\begin{equation}
  B = A \frac{6}{13 \pi^2} \frac{\overline{m}_\mu^2 (T)}{T^2} L^{\star}_e,
\label{eq:Bprediction}
\end{equation}
where $A \simeq 1$ \cite{misha} and
\begin{equation}
  \frac{\overline{m}_\mu^2 (T)}{T^2} = \frac{1}{6} f_\mu^2 +
  \frac{1}{3} f_\mu^2 \left( \frac{v(T)}{T} \right)^2
  \leq \frac{1}{2} f_\mu^2 = 1.8 \times 10^{-7}.
\end{equation}
The resulting baryon-to-photon ratio in this case is
\begin{equation}
  \eta = (1.8 \pm 1.68) \times 10^{-10}
\end{equation}
This should be compared to the
preferred value from BBN, e.g. \cite{KKS2} $\eta =
(4.0^{+1.5}_{-0.9})\times 10^{-10}$. Thus
we find agreement with the required value 
at the upper edge of the 95\% CL region!.

Notice that, if the electroweak phase transition is strongly
first order with $\langle
\phi(T)\rangle/T$ larger than unity after the transition, the sphaleron
processes are frozen-out and absent after the transition.
In this case the chemical
equilibrium before the transition determines the baryon asymmetry.
The only flavor-dependent effects before the transition are the
Yukawa interactions of electrons and muons with the uncondensed
Higgs boson.  In the case of vanishing total lepton asymmetry
$L_e + L_\mu=0$, we now find
\begin{equation}
  \frac{\overline{m}_\mu^2 (T)}{T^2} = \frac{1}{6} f_\mu^2
  = \frac{\pi \alpha_W}{3} \frac{m_\mu^2 (0)}{m_W^2} = 6.0 \times 10^{-8}.
\end{equation}
This translates into baryon and lepton to entropy ratios of
\begin{equation}
  B = \frac{6}{13\pi^2} (6.0 \times10^{-8})
  L_e^{\star} = (8.6 \pm 8.0) \times 10^{-12}.
\end{equation}
This corresponds to a current baryon-to-photon ratio of
$\eta = (6.0 \pm 5.6) \times 10^{-11}$, which is off by more than a
factor of three.

\section{Model Building}

We have seen that the preferred value of $L_e$ from Eqn.~(\ref{eq:inputLS}),
$L_e^\star=2 L_e^{{\rm NUC}}$, 
together with the relation $L_e = - L_\mu$ gives the correct order of
magnitude and sign for the baryon asymmetry.  We find
this a remarkable coincidence. 

Suppose however one takes the preferred value of the
lepton asymmetry to be $L_e^{{\rm NUC}}$, {\it i.e.} let us not 
allow any room for extra degrees of freedom during nucleosynthesis.
Then from Eqn.~(\ref{eq:Bprediction}) the baryon asymmetry turns
out to be correct except for a factor of two or so.  A natural
question then is if there are corrections that can fix this
factor-of-two discrepancy so that the generation of
the observed small baryon asymmetry from the magnitude of the lepton 
asymmetry currently preferred from the BBN is a
realistic possibility.  We find that there
are many ways to achieve this.  Another natural question is if there
is an appropriate leptogenesis mechanism which can create a large
lepton asymmetry with $L_e = - L_\mu$ in a simple way.

The simplest possibility to enhance the baryon asymmetry is to consider
the MSSM where all sleptons and squarks are heavier
than the electroweak phase transition temperature while the
entire Higgs sector, $h^0$,
$H^0$, $A^0$ and $H^\pm$ is light.  In the approximation where one ignores
their masses, the lepton doublets interact only with the $H_d$ doublet
with the Yukawa coupling $f_l / \cos \beta$.  Here $\tan \beta \equiv
\langle H_u \rangle/\langle H_d \rangle$ is the vacuum angle.  In this
limit, the only change from the case of the SM is to replace the
Yukawa couplings $f_l$ by $f_l/\cos\beta$, which enhances the plasma
mass effects.  The net result is an enhanced baryon asymmetry which
brings the predicted value into the required range for a moderate
value of $\tan\beta$.  If the masses of the Higgs bosons cannot be
neglected, the enhancement effect is reduced.  But it is clear that a
realistic value of the baryon asymmetry can be easily achieved.

There are many other possible enhancement mechanisms of the baryon
asymmetry.  For instance, light higgsinos and sleptons also
contribute to the plasma mass of the lepton doublets.  These effects
are enhanced by $1/\cos^2 \beta$ but Boltzmann-suppressed by their masses $\sim
e^{-m/T}$.  For suitable values of $\tan\beta$
and the slepton and higgsino masses
estimates indicate that the required factor of two is generated. 
(A detailed quantitative analysis involves the generalization
of the formulae in \cite{IIMS} to include the individual
lepton asymmetries.)  It is therefore clear that there are quite simple
extensions of the SM which fairly naturally provide the required
factor of two.

We now turn to the question of whether it is possible to generate
a large lepton asymmetry with $L_e=-L_\mu$ in a natural and elegant 
fashion.
Such leptogenesis with $L_e = -L_\mu$ can be achieved naturally
by utilizing the Affleck--Dine mechanism \cite{AD}.\footnote{It was discussed 
recently also in \cite{McDonald} that one can generate a large lepton 
asymmetry by the Affleck--Dine mechanism.  The author however required 
an even larger asymmetry than what we discuss to keep the electromagnetism 
as well as sphalerons out of equilibrium to solve the monopole problem 
and avoid the overproduction of baryon asymmetry 
\cite{BBD,Linde,DMO,RS,bo}.  Therefore the aim of the paper is orthogonal 
to ours.}  This requires the operator \cite{CGMO}
\begin{equation}
  \int d^4 \theta (m_{3/2} \theta^2) (m_{3/2} \bar{\theta}^2)
  \frac{L_e^* L_\mu H_u^* H_u}{M_X^2},
\end{equation}
where the supersymmetry-breaking spurions are inserted.  This operator
preserves the total lepton number, while breaking $L_e$ and $L_\mu$
individually.  The energy scale of this operator, $M_X$, can be, for example,
the (reduced) Planck scale $M_*= 2\times 10^{18}~{\rm GeV}$.
The $D$-flat direction $|L_e|^2 + |L_\mu|^2
= |H_u|^2$ is lifted by this operator and the field acquires a large
``angular momentum'' in the internal space.  This corresponds to the 
generation of individual lepton numbers satisfying $L_e=-L_\mu$.
This leads to an estimate of the lepton number,
\begin{equation}
  L_e = -L_\mu \simeq \frac{\phi_0^4 T_{RH}}{m_{3/2} M_X^2 M_*^2},
\label{eq:asym}
\end{equation}
where $T_{RH}$ is the reheating temperature of primordial inflation,
$m_{3/2}$ is the typical mass of the sleptons and Higgs bosons, and $\phi_0$
is the initial amplitude of the slepton expectation values.
Even taking account of the
constraint from gravitino overproduction $T_{RH} \lesssim
10^9$~GeV\footnote{This bound is obtained considering the 
thermal production of gravitinos. However, it has been
recently pointed out  that  this mechanism of productio is  overcome by the non-thermal generation of gravitinos \cite{nontherm}.}
and $m_{3/2} \sim 1$~TeV, the initial value of the amplitude
can be relatively small $\phi_0 \gtrsim (10^{-3} L_e M_X M_*)^{1/2}$.
Taking $M_X\sim M_*$, Eqn.~(\ref{eq:asym})
shows that $\phi_0 \sim 10^{15}$~GeV is
sufficient to generate the large lepton asymmetry that we require. 
Note that the detailed mechanism for generating a large initial amplitude,
$\phi_0$, is model-dependent; it could be a negative mass-squared during the
inflationary epoch \cite{DRT,dterm} or quantum
effects \cite{GMO}.\footnote{Since the total lepton number is
preferably conserved within our scenario, the neutrino masses should
be Dirac rather than Majorana.  The atmospheric neutrino
oscillation prefers a small Yukawa coupling of order $h_\nu \lesssim
10^{-12}$.  Even though this Yukawa coupling lifts our flat
direction, a negative mass squared of, for instance, $-H_{inf}^2$
during inflation, generates an initial amplitude of $\phi_0 \sim
H_{inf}/h_\nu$ which is well beyond what we need given the typical
value of the Hubble constant during inflation
$H_{inf} \sim 10^{11}$--$10^{13}$~GeV.  Such a small Yukawa coupling 
could be a natural consequence of a flavor symmetry \cite{Carone}.}

One final concern is if this scenario is consistent with the reported
atmospheric neutrino oscillation: If the generated asymmetry $L_\mu$ is
converted partially to an asymmetry in $L_3$, it could then generate too
large a baryon asymmetry because of the large tau Yukawa coupling
$f_\tau$.  This fortunately does not happen.  By the time of the
electroweak phase transition, the probability for neutrino
oscillation is suppressed by $\sin^2
(\Delta m^2 t / 4 E_\nu)$, where $t \sim M_*/m_W^2$ and $E_\nu \sim
m_W$.  Substituting the relevant $\Delta m^2$ into this expression
then shows that the oscillation to $L_3$ is negligible.

\section{Conclusion}

Over the years, many mechanisms for the generation of the 
tiny observed baryon asymmetry have been proposed and we have
very little idea which if any is the correct one.  Furthermore, many
of the proposed baryogenesis mechanisms are not able to predict the 
resulting baryon asymmetry to better than an order of magnitude
(sometimes many).  On the other hand,
so far there is no observational evidence excluding the possibility
that the lepton asymmetry in the Universe is almost as large as the
present entropy density.  On the contrary, the current measurements
of the light element abundances may prefer such an asymmetry to
reconcile BBN theory with observations.  In this paper, we have
made a simple observation which seems quite surprising to us: If the
asymmetries in electron number and muon number are equal and
opposite and of the size indicated by nucleosynthesis 
considerations, a baryon asymmetry of the observed size is
naturally generated within the Standard Model itself due to the
small but non-zero muon Yukawa coupling.  This might just be a
coincidence, but it is quite an intriguing one!

\section*{Acknowledgements} 

AR thanks M. Shaposhnikov for useful
conversations. JMR and HM thank the Aspen Center for Physics where a part of
this work was done.  HM also thanks the Institute for Nuclear Theory at the University of 
Washington for its hospitality and the Department of Energy for partial 
support during the completion of this work and Jose Valle and Raymond Volkas 
for useful discussions.  HM was supported in part by the
U.S. Department of Energy under
Contracts DE-AC03-76SF00098, in part by the National Science
Foundation under grant PHY-95-14797.  Both JMR and HM were supported by 
the Alfred P. Sloan Foundation.

\end{document}